\documentclass[final]
  {aipproc}

\layoutstyle{6x9}

\usepackage{graphicx}
\usepackage{epsfig}
\begin{document}

\newcommand{\be}[1]{\begin{equation}\label{#1}}
 \newcommand{\ee}{\end{equation}}

\title{The exact dynamical solution for two dust shells collapsing towards a black hole}


\author{Shuang Nan Zhang}{
  address={Physics Department and Center for Astrophysics,
Tsinghua University, Beijing, 100084, China\footnote{email:
zhangsn@tsinghua.edu.cn}} }

\author{Yuan Liu}{}

\begin{abstract}
 The gravitational collapse of a star is an important issue both for
general relativity and astrophysics, which is related to the well
known ``frozen star" paradox. Following the seminal work of
Oppenheimer and Schneider (1939), we present the exact solution for
 two dust shells collapsing towards a pre-existing black hole. We find that the inner region of the shell is
influenced by the property of the shell, which is contrary to the
result in Newtonian theory and and the clock inside the shell
becomes slower as the shell collapses towards the pre-existing black
hole. This result in principle may be tested experimentally if a
beam of light travels across the shell. We conclude that the concept
of the ``frozen star" should be abandoned, since matter can indeed
cross a black hole's horizon according to the clock of an external
observer. Since matter will not accumulate around the event horizon
of a black hole, we predict that only gravitational wave radiation
can be produced in the final stage of the merging process of two
coalescing black holes. Our results also indicate that for the clock
of an external observer, matter, after crossing the event horizon,
will never arrive at the ``singularity" (i.e. the exact center of
the black hole).
\end{abstract}

\maketitle


\section{Introduction}

  The ``frozen star" paradox is a well known novel phenomenon predicted by
general relativity, i.e. a distant observer ($O$) sees a test particle falling towards
a black hole moving slower and slower, becoming darker and darker, and is eventually
frozen near the event horizon of the black hole. However, as we discussed in \cite{20},
several possible explanations to this phenomenon have been proposed, but none of these
is completely satisfactory. In the previous work, we found that an external observer
should be able to observe matter falling in a black hole, based on our {\it stationary}
solution of a dust shell around a pre-existing black hole. In this paper we further
obtain the exact {\it dynamical} solution for two shells collapsing towards a
pre-existing black hole, in particular to find the difference between this solution and
that for the gravitational collapse of a uniform dust ball \cite{14}, and the
implication for the ``frozen star" paradox.

\section{The dynamical solution}
We follow the method adopted in the seminal work of Oppenheimer and
Schneider (1939) \cite{14}, i.e. obtaining the solution in the
comoving coordinates first and then transforming it into the
ordinary
 coordinates.

The general form of the metric in the comoving coordinates is ($G =
c = 1$ is adopted throughout this paper)
 \be{eq15} ds^2  = d\tau ^2
- e^{\bar \omega } dR^2  - e^\omega (d\theta ^2  + \sin ^2 \theta
d\phi ^2 ), \ee where  $\bar \omega (R,\tau )$
 and  $\omega (R,\tau )$
 are the unknown functions. In the
comoving coordinates, the only non-zero component of the energy
momentum tensor is  $T_4^4  = \rho $. As the results in \cite{14},
we can obtain the following equations from the field equation
\be{eq16}e^\omega   = (F\tau  + G)^{4/3},\ee
 \be{eq17}e^{\bar \omega }  =
e^\omega  \omega '^2 /4, \ee
 \be{eq18}8\pi \rho  = \frac{4}{3}\;(\tau
+ G/F)^{ - 1} (\tau  + G'/F')^{ - 1},\ee where $F$
  and  $G$ are
arbitrary functions of $R$. Dot and prime are the partial
differentiation with respect to $\tau $
  and  $R$, respectively. At  $\tau=0$, we
obtain \be{eq19}F =  - \sqrt {6\pi \rho _0 (G^2  + C_1 )}, \ee where
$C_1 $ is the integration constant. As in \cite{14}, we choose $G =
R^{3/2}$.

As obtained in \cite{14}, the value of $F$ in region V is $ -
\frac{3}{2}\sqrt {r_0 } $, where  $r_0  = 2M$ ($M$ is the total mass
of the system). Then to assure the solutions are continuous at the
boundaries, the results of $F$ in other regions should be \be{} F =
\left\{ {\begin{array}{*{20}c}
   { - \frac{3}{2}\sqrt {r_0 '} \;\;\;\;\;\;\;\;\;\;;R < a_1 '}  \\
   { - \frac{3}{2}\sqrt {\frac{{8\pi \rho _1 }}{3}(R^3  - a_1 ^3 ) + r_0 ^{\prime \prime } } \;\;\;;a_1 ' < R < a_1 }  \\
   { - \frac{3}{2}\sqrt {r_0 ^{\prime \prime } } \;\;\;;a_1  < R < a_2 '}  \\
   {\begin{array}{*{20}c}
   { - \frac{3}{2}\sqrt {\frac{{8\pi \rho _2 }}{3}(R^3  - a_2 ^3 ) + r_0 } \;\;\;;a_2 ' < R < a_2 }  \\
\end{array}}  \\
\end{array}} \right.,
\ee where  $r_0  = 2M,\;r_0 ' = 2m$, $r_0 ^{\prime \prime }  = 2(m +
m_1 )$.

Up to now we have obtained the solution in the comoving coordinates.
Then we try to transform the solution into the ordinary coordinates,
in which the metric has the form as \be{eq20} ds^2  = B(r,t)dt^2  -
A(r,t)dr^2  - r^2 (d\theta ^2  + \sin ^2 \theta d\phi ^2 ). \ee

The transformation of $r$ is obvious. We must choose $r=e^{\omega
/2} = (F\tau  + G)^{2/3}$.

Using the contravarient form of the metric and requiring that the
$g^{tr}$ term vanishes, we have
\be{eq22} t'/\dot t = \dot rr'. \ee

Substituting the expressions of $r$ in five regions into Eq.
(\ref{eq22}), we can obtain five partial differential equations of
$t(R,\tau )$.

In region V, the solution of Eq. (\ref{eq22}) is \be{eq23} t =
L(x),\;x = \frac{2}{{3\sqrt {r_0 } }}(R^{3/2}  - r^{3/2} ) - 2\sqrt
{rr_0 } + r_0 \ln \frac{{\sqrt r  + \sqrt {r_0 } }}{{\sqrt r  -
\sqrt {r_0 } }}, \ee where $L$ is an arbitrary function of $x$.
Since the metric in the ordinary coordinates in region V is the
Schwarzschild metric, $L(x)$ should be $x$.

In general case, the partial differential equations in other regions
can be solved numerically only, using the solution in region V as
the boundary condition.

After obtain the transformations in every region, we could know how
the shells evolve with the coordinate time. Fig. 1a and b show the
configuration of this problem and the evolution of the shells in the
ordinary coordinates. In Fig. 1c, we show the evolution curves of
the outer boundary of shell 1 in the case with or without shell 2.
The outer boundary of shell 1 will take more time to approach the
asymptotic line in the case with shell 2.

\begin{figure}
\includegraphics[width=0.3\textwidth]{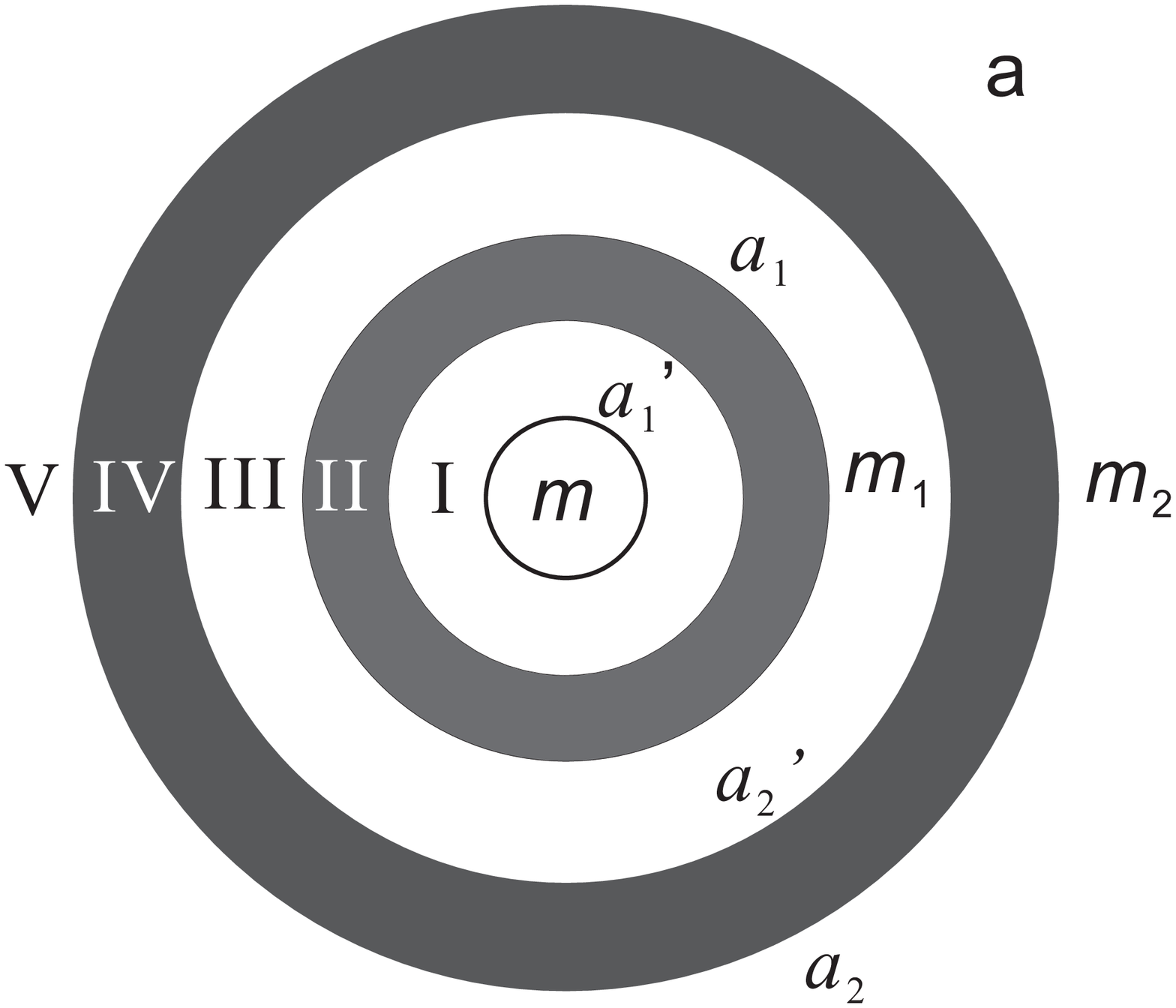}
\hspace{0.1cm}
\includegraphics[width=0.33\textwidth]{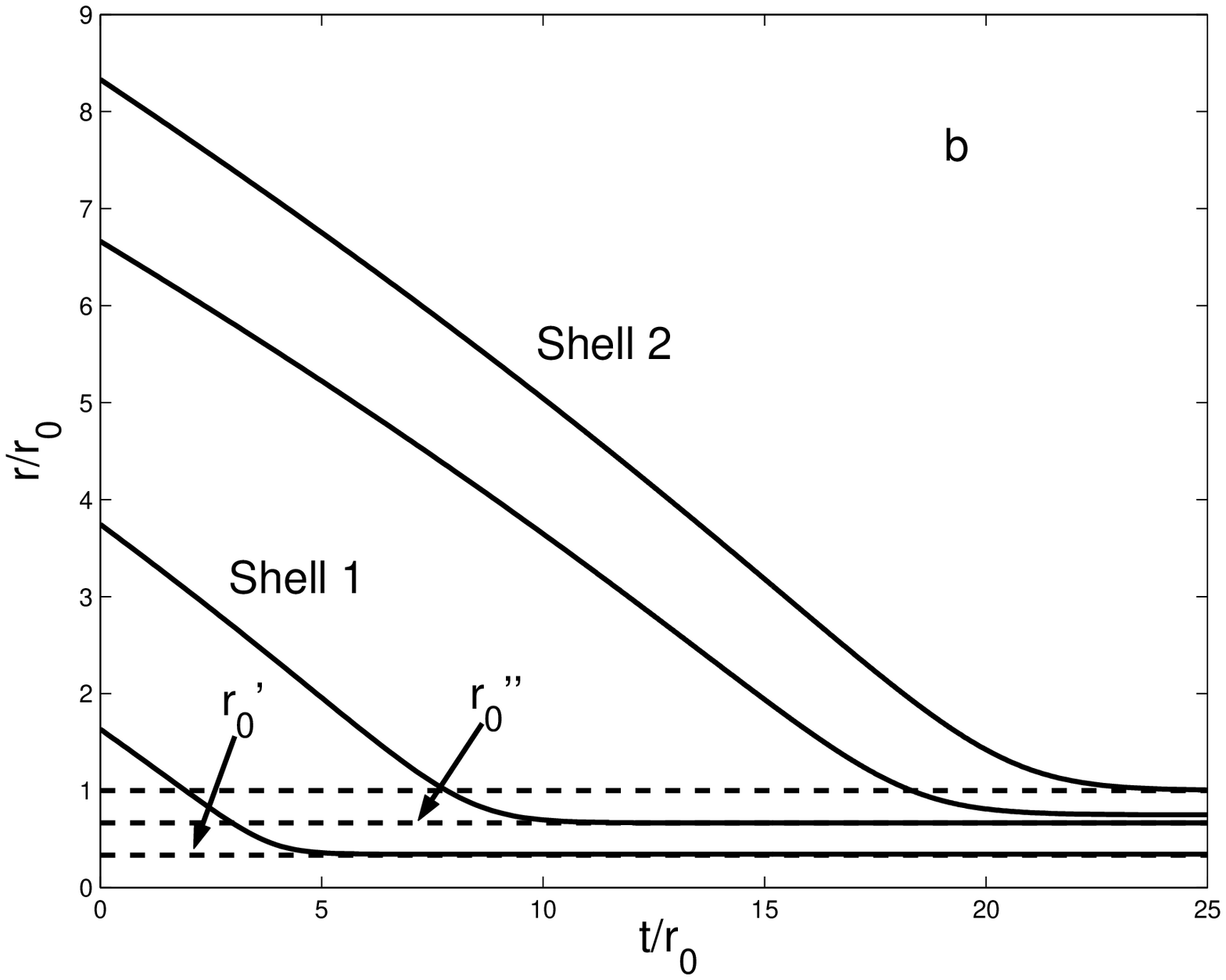}
\hspace{0.1cm}
\includegraphics[width=0.33\textwidth]{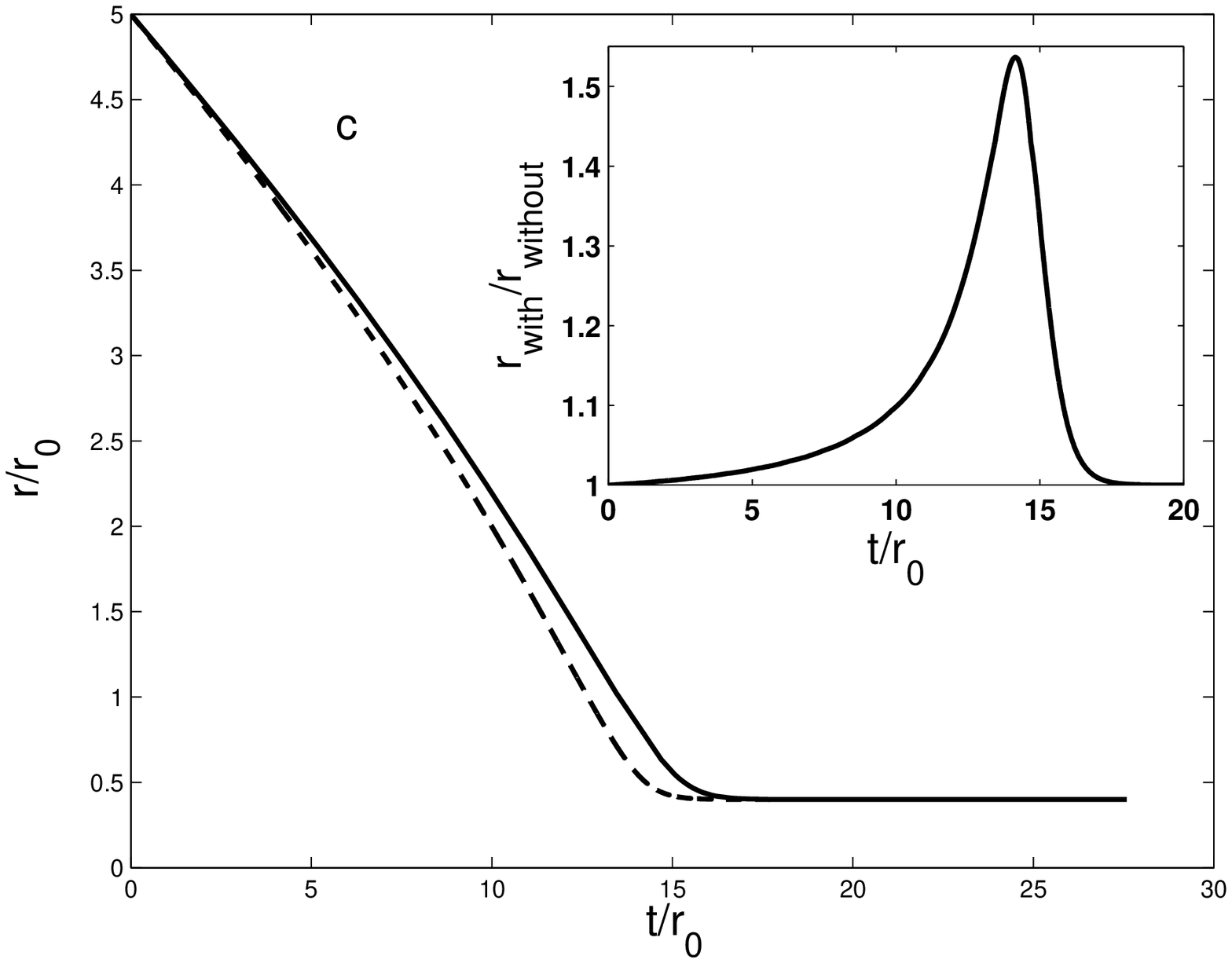}
\caption{\label{fig6}(a) is  the configuration of the dynamical
solution for two shells in comoving coordinates.  $m$,   $m_1 $ and
$m_2 $ are the masses of the black hole, the shell 1 and 2,
respectively. $a_1 '$
  and  $a_1 $
 are the radii of the
inner and outer boundaries of shell 1, respectively.  $a_2 '$
 and  $a_2$
 are the radii of the
inner and outer boundaries of shell 2, respectively. (b) is
evolution curves of the dynamical solutions for $a_2 = 10r_0 $, $a_2
' = 8r_0 $, $a_1 = 5r_0 $, $a_1 ' = 2.5r_0 $, $r_0 ' = 1/3r_0 $, and
$r_0 '' = 2/3r_0 $ with coordinate time. (c) is the comparison of
the evolution of the outer boundary of shell 1 between the case with
(solid) or without (dashed) shell 2. The parameters of the shells
are  $a_2 = 10r_0 $, $a_2 ' = 6r_0 $, $a_1 ' = 2.5r_0 $, $a_1 = 5r_0
$, $r_0 ' = 1/5r_0 $, and  $r_0 '' = 2/5r_0 $, where $r_0$ is the
Schwarzschild radius corresponding to the total mass of the system,
i.e., the sum of the masses of the pre-existing black hole, shell 1
and shell 2. The inset is the ratio of the solid line to the dashed
line. It is seen clearly that outside matter, though spherically
symmetric, does influence the motion of matter inside.}

\end{figure}

\section{Discussions and Conclusions}

In the Newtonian gravitation theory, if the matter is spherically
symmetric, the outer matter will not influence the inner region.
However, as shown in Sec. II, in general relativity case, the clock
in the inner region of the shell is slower compared with the case
without the shell. This effect may be testable experimentally in
principle, e.g. if a beam of light travels across the shell and a
parallel beam of light passes outside the shell (far enough from the
shell), then the two beams of light will travel at different
velocities with respect to the observers outside the shell.

The inner shell can  cross the Schwarzschild radius in the two shell
case. In this sense we could observe the matter falls into a black
hole and the ``frozen star" paradox discussed in \cite{10} is
naturally solved. As pointed out previously by us, the origin of the
``frozen star" is the ``test particle" assumption for the infalling
matter, which neglects the influence of the infalling matter itself
on the metric  \cite{20}. In fact, for some practical astrophysical
settings, the time taken for matter falling into a black hole is
quite short, even for the external observer \cite{20}. We find that
the infalling matter will not accumulate outside the event horizon,
and thus the quantum radiation and Gamma Ray bursts predicted in
\cite{11} and \cite{12} are not likely to be generated. We predict
that only gravitational wave radiation can be produced in the final
stage of the merging process of two coalescing black holes. Future
simultaneous observations by X-ray telescopes and gravitational wave
telescopes shall be able to verify our prediction.

It is also interesting to note that, as can be seen from Fig.
\ref{fig6}b, in ordinary coordinates, the matter will not collapse
to the singularity  $(r = 0)$
 even with infinite coordinate time (if $r_0' = 0$, the inner boundary of the inner shell will take infinite time
  to arrive at  $r = 0$). It means that in real astrophysics sense, matter can never
arrive at the singularity (i.e., the exact center of the black hole)
with respect to the clock of the external observer. Therefore, no
gravitational singularity exists physically, even within the
framework of the classical general relativity.


\begin{theacknowledgments}
 We thank many discussions with Sumin Tang, Richard Lieu, Kinwah Wu, Kazuo Makishima,
Neil Gehrels, Masaruare Shibata, Ramesh Narayan, Zheng Zhao, Zonghong Zhu and Chongming
Xu. SNZ is grateful to Prof. Sandip Chakrabarti for his great effort in organizing this
conference, and to the great hospitality of S.N. Bose National Centre for Basic
Sciences, Kolkata, India. Many participants of this conference, especially Prof. Roy
Kerr, are greatly appreciated for interesting discussions. SNZ acknowledges partial
funding support by the Yangtze Endowment from the Ministry of Education at Tsinghua
University, Directional Research Project of the Chinese Academy of Sciences under
project No. KJCX2-YW-T03 and by the National Natural Science Foundation of China under
project no. 10521001, 10733010 and 10725313.
\end{theacknowledgments}



\bibliographystyle{aipproc}   

\bibliography{sample}

\IfFileExists{\jobname.bbl}{}
 {\typeout{}
  \typeout{******************************************}
  \typeout{** Please run "bibtex \jobname" to optain}
  \typeout{** the bibliography and then re-run LaTeX}
  \typeout{** twice to fix the references!}
  \typeout{******************************************}
  \typeout{}
 }


\end{document}